\documentclass[aps,twocolumn,showpacs,floatfix,prd]{revtex4-1}
\usepackage{bm,enumerate,dcolumn,graphicx,color,amsmath,amssymb}
\begin{document}

\title{Stimulated deceleration of diatomic molecules on multiple rovibrational transitions with coherent pulse trains}

\author{Ekaterina Ilinova}
\affiliation{Department of Physics, University of Nevada, Reno,
Nevada 89557, USA}
\author{Jonathan Weinstein}
\affiliation{Department of Physics, University of Nevada, Reno,
Nevada 89557, USA}
\author{Andrei Derevianko}
\affiliation{Department of Physics, University of Nevada, Reno,
Nevada 89557, USA}

\date{\today}

\begin{abstract}
We propose  a method of stimulated laser decelerating of  diatomic molecules by counter-propagating $\pi$-trains of ultrashort laser pulses.
The decelerating cycles occur on the rovibrational transitions inside the same ground electronic manifold, thus avoiding the common problem of radiative branching in Doppler cooling of molecules. By matching the frequency comb spectrum of the pulse trains to spectrum of the R-branch rovibrational transitions we show that stimulated cooling can be carried out on several rovibrational transitions simultaneously, thereby increasing number of cooled molecules.  The exerted optical force does not rely on the decay rates in a system and can be  orders of magnitude larger than the typical values of scattering force obtained in conventional Doppler laser cooling schemes.
\end{abstract}


\maketitle 


While laser cooling is one of the key techniques of modern atomic physics~\cite{MinLet87Book,MetStr99Book, BerMal10_Book},
neutral molecules are notoriously challenging to cool to ultracold temperatures.
To accomplish this feat  one must exercise precise control over a multitude of internal degrees of freedom.
A breakthrough in molecular cooling techniques --- even for diatomic molecules --- is anticipated to enable substantial progress in quantum information processing~\cite{DeM02}, matter-wave interferometric
sensors~\cite{DutJaaMey05},
quantum-controlled chemical reactions~\cite{BalDal01} and precision measurements~\cite{KozLab95, BarMarRyc02,BodGiaDal02}.

To date, the coldest samples of diatomic molecules were obtained  by  assembling them from ultracold atoms via photo- or magneto- association. This approach produced a gas of ground-state polar molecules near quantum degeneracy~\cite{NiOspMir08,HecNagGri04}. However, so far only bialkali molecules
have been produced in this way, and the number of molecules produced is fairly small ($\sim 10^4$).
Deceleration of the  ensembles of cold molecules in electrostatic Stark decelerators has been demonstrated \cite{BetFloBer00,BetHenRoi02}.
Recently there has been demonstrated the Sisyphus cooling of polyatomic molecules \cite{ZepEngGlo12}. Direct laser cooling of molecules could yield substantially  larger samples for a  wider range of  species.

In traditional Doppler cooling the radiative force originates from momentum transfer to atoms from a laser field and subsequent spontaneous emission in random directions.
Repeating this optical cycle tens of thousands of times  can slow down thermal beams and cool atomic samples down to the Doppler limit (typically mK). Unfortunately, most atoms and all molecules  can radiatively decay to a multitude of states. Exciting population from all
these lower-energy states requires a large number of lasers, which makes the conventional scheme impractical. Only for a narrow class of molecules with highly-diagonal Franck-Condon overlaps (i.e., ``near-cycling'' electronic transitions), this branching problem can be mitigated~\cite{StuSawWangYe08,ShuBarGle09,BarShuNor12}.

Here we explore an alternative:  employ  absorption and  {\em stimulated} emission on weak transitions inside the same molecular electronic manifold.
%
The concept of ``stimulated cooling" was put forward by Kazantsev~\cite{Kaz74} in 1974; he proposed illuminating a two-level atom with a sequence of alternatively directed and oppositely detuned $\pi$ pulses. For an atom initially in the ground state, a  $\pi$-pulse impinging from the left would drive the population upward, while the $\pi$-pulse from the right cycles the population back to the ground state. The net change of atomic momentum is twice the photon recoil momentum $p_r=\hbar k_c$, where $k_c$ is the laser wave-vector.
The advantage of this scheme over Doppler cooling is that the momentum transfer can occur much faster than the radiative decay; this is crucial for the use of rovibrational transitions, for which Doppler cooling is impractical due to the very long radiative lifetimes.
Stimulated cooling with counter-propagating $\pi$-pulses was explored in a number of works~\cite{NebSzo74,FriWil76,PalLam86}.

Closely related to the stimulated cooling by $\pi$-pulses is bichromatic force cooling~\cite{CasMet01,GriOvcSid90,GriSOvc94,ParMiaBoc04,SGriOvc97,VoiDanNeg88,VoiDanNeg94,YatMet04}. Here the trains of Kazantsev's  $\pi$-pulses effectively arise from the beating of
two  counter-propagating CW lasers of  different frequencies. While the experiments so far
have been limited to atoms, bichromatic force cooling of diatomic molecules on near-cycling electronic transitions was recently proposed~\cite{ChiEyl11}. Notice that the term ``cooling'' is used loosely in the context of stimulated processes in the literature: compared to spontaneous emission, stimulated emission does not lead to entropy reduction and thus to the phase-space compression. Considering the existing practice, we will use the terms stimulated ``cooling'' and ``slowing'' interchangeably.  

In this work we explore molecular slowing with coherent pulse trains.  Pulse trains offer several advantages.  First, the broadband structure of mode-locked lasers allows one to address multiple rotational levels simultaneously.  Secondly, it simplifies stabilization of the multiple frequencies involved, even for species for which building reference cells is impractical \cite{RevModPhys.75.325}.  Finally, by manipulating the phases of successive pulse trains, as shown below, the capture range of velocities addressed can be tuned. Looking at the state-of-the-art frequency combs (FC) \cite{SchHarYos08,AdlCosTho09, LeiMarBye11,VodSorSor11}, we note that a fiber-laser-based FC with 10~W average power was demonstrated~\cite{SchHarYos08} and the authors argue that the technology is scalable above 10 kW average power.  The spectral coverage was expanded from optical frequencies to ultraviolet and to mid-IR spectral regions~\cite{AdlCosTho09, LeiMarBye11,VodSorSor11}. The high resolution quantum control via the combination of pulse shaping and frequency comb was shown  in \cite{Ye07, JiaHuaLea07,StoCruFla06,StoMatAvi08}. The experiments on line-by-line addressing have been done \cite{JiaHuaLea07}.
In  \cite{JiaHuaLea07} the authors have demonstrated the high-fidelity programmable, and line-by-line manipulation of the amplitude and phase of more than 100 individual comb components (with 5 GHz teeth spacing) in a spectral span of about 0.5 THz.

We focus on the transitions between the rovibrational levels $(\upsilon,J)$ {\em inside} the ground electronic potential.  The frequencies $\nu_{J,J+1}^{\upsilon,\upsilon'}$  of the $R$-branch ($J \rightarrow J'=J+1$) electric-dipole-allowed transitions between  rovibrational manifolds read
\begin{equation}
\nu^{v,v'}_{J,J+1} \approx \nu^{v,v'}_{0,1}+(3B_{v'} - B_v)  J + (B_{v'} - B_v) J^2 \, ,\label{Eq:Rbranchfreq}
\end{equation}
where $\nu^{\upsilon,\upsilon'}_{0,1}$ is the frequency of the $X(v,J\!=\!0) \rightarrow X(v',J'\!=\!1)$ transition and
$B_{\upsilon}$ are the rotational constants for the $\upsilon$-th vibrational level~\cite{Herz1950}.
This pattern matches the frequency comb (FC) spectrum which consists of
a series of sharp equidistant peaks (teeth) located at
\begin{equation}
\nu_n = \nu_c +\nu_\mathrm{rep}\times  n -  \frac{\Phi}{2\pi T} \, ,
\label{Eq:CombSpectrum}
\end{equation}
where $n$ is an integer number, $\nu_c$ is the carrier frequency,
the teeth-spacing $\nu_\mathrm{rep}=1/T$ is defined in terms of pulse repetition period $T$, and $\Phi$ is the carrier-envelope-offset (CEO) phase. We will focus on the $\upsilon=0\rightarrow \upsilon'=1$ transitions.  The molecular and the comb spectra could be matched by choosing
$\nu_c=\nu^{0,1}_{0,1}$ and $\nu_\mathrm{rep}=(3B_{1}-B_0)/n'$, with $n'$ being an integer. The R-branch spectrum~(\ref{Eq:Rbranchfreq}) becomes Doppler-shifted  for moving molecules;
we will adjust $\Phi$ to follow the Doppler shift of the band-head frequency.

As an example, consider the LiCl molecule; here $\nu_c\approx 19.29$ THz with  $T= 0.95$ ns for the number of teeth between nearby rovibrational transitions $n'=40$. For the PbO molecule, $\nu_c \approx 21.63 $ THz with  $T=0.98$ ns for  $n'=15$.

\begin{figure}[h]
\begin{center}
\includegraphics*[width=3in]{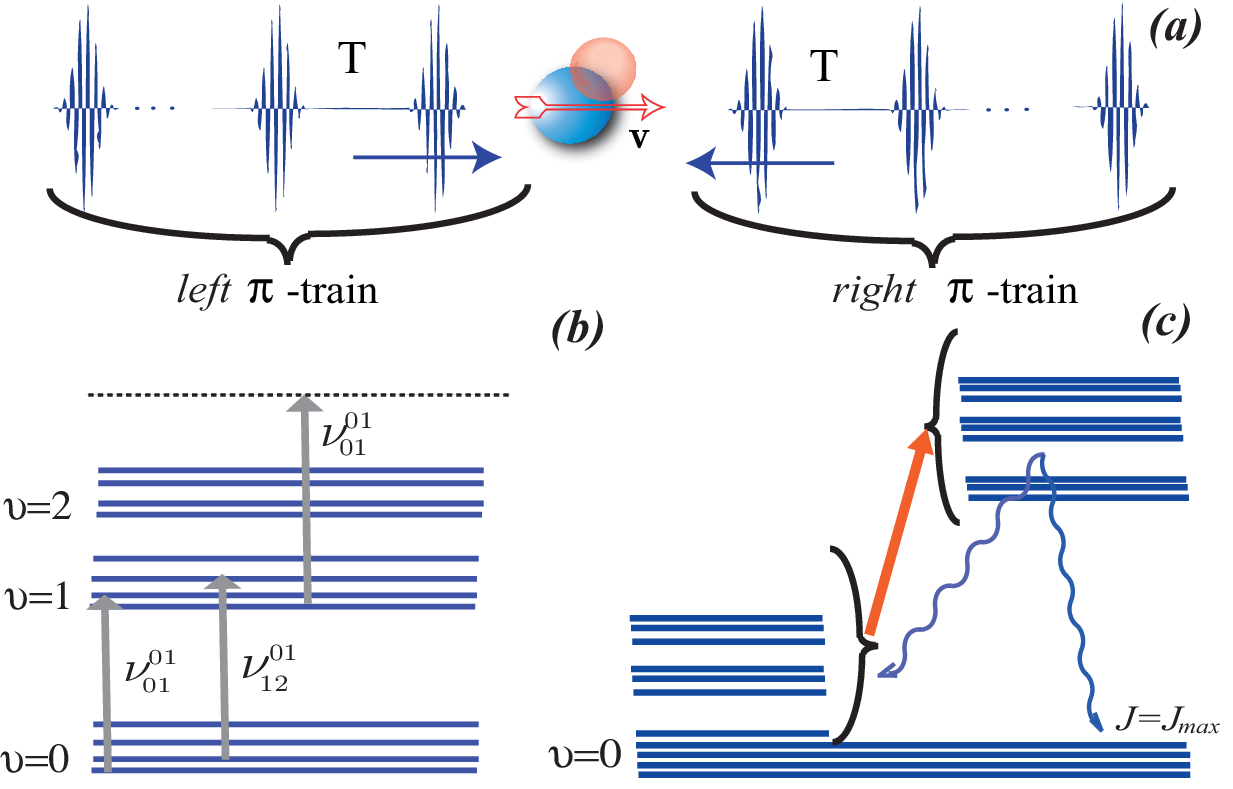}
\end{center}
\caption
{(Color online) (a) Stimulated slowing by $\pi$-trains. (b) R-branch rotational transitions form a spectrum that matches the frequency comb spectrum of pulse trains.
 (c)  Reseting molecular population to the ground vibrational manifold $X(\upsilon=0, J)$ through the exited electronic state $A$ with broad-band laser pulses.
\label{Fig:Setup}}
\end{figure}

Based on these observations we propose the following scheme of stimulated cooling of molecules with pulse trains.
As illustrated in Fig.~\ref{Fig:Setup}(a), the essential idea is to replace each of the Kazantsev's 
 $\pi$-pulses by a train of $N$ pulses, each of pulse area $\theta=\pi/N$.
Interference of molecular probability amplitudes induced by the pulses leads to a frequency-dependence of the cooling force resembling the FC spectrum. Cooling will occur simultaneously on several rovibrational transitions.
The elementary cooling sequence (or cycle) will consist of two subsequent counter-propagating $\pi$-trains.  Each cycle will transfer twice the recoil momentum $p_r=\hbar k_c$.
To maximize the optical force we assume that the time delay between  trains is negligible.

 For a moving molecule, Doppler shifts due to the co- and counter-propagating trains have opposite signs.
 The counter-propagating/upward-stimulating train must be tuned below the resonance and the co-propagating train above the resonance. This can be attained by keeping the carrier frequencies of the two trains the same but having two different CEO phases.

Before proceeding to the full analysis, note that the spectrum given in Eq. (\ref{Eq:Rbranchfreq}) is for the R-branch.  To avoid driving P-branch transitions, one would need to eliminate the low-frequency ($n<0$) half of the comb spectrum in Eq.~(\ref{Eq:CombSpectrum}), thus reducing the problem to a collection of (separate) two-level systems.  This truncation may be accomplished by  installing a dispersive element and a static mask at the output of the cavity; the mask  would cut out a predefined spectral window.  Such a technique has been used by the Orsay group~\cite{VitChoAll08} in pulsed laser cooling of Cs$_2$ molecules.

Furthermore, we would like to avoid driving the $v=1 \rightarrow v=2$ transitions, see Fig.~\ref{Fig:Setup}(b). Fortunately, due to anharmonicity, $\nu^{1,2}_{0,1} < \nu^{0,1}_{0,1}$, so the $\nu^{1,2}_{1,2}$ bandhead is shifted to lower frequencies, already masked-out. Thereby only the higher-$J$ rotational transitions will overlap with the truncated comb spectrum. The maximum $J$ of the $v=1\rightarrow 2$ band overlapping with the comb spectrum could be estimated as $J_{\max} =(\nu^{1,2}_{0,1}-\nu^{0,1}_{0,1})/(B_2+B_1) \approx \omega_e x_e/B_e$, where we used the conventional spectroscopic notation~\cite{Herz1950}. For LiCl and PbO, this limits us to $J<6$ and $J<10$ correspondingly.  This limitation to low $J$  is consistent with our neglect of higher-order non-rigid-rotor terms in Eq.~(\ref{Eq:Rbranchfreq}). This limitation is of little consequence: at temperatures below 300 K, the majority of the molecular population is in these low-$J$ levels.
Another peculiarity is that  dipole matrix elements  vary across rovibrational transitions, while the trains must execute the full $\pi$ rotation for all transitions simultaneously; to meet this requirement pulse shapers would need to be used so that the Rabi frequency of various transitions remains the same.

{\em Formalism. --} Since the FC fields couple only pairs of molecular levels, the problem is
reduced to finding the time evolution and optical force on a two-level system.
Due to the rapid  nanosecond time-scale of $\pi$-train cycles we may neglect  slow radiative decay of the vibrational levels (relevant lifetimes are on the order of a millisecond).
We also approximate short non-overlapping laser pulses by $\delta$-functions.
In this limit, the  pulses are fully characterized by the pulse area
$\theta$ and phase $\Phi$.

In the interaction picture, the propagator
$U_\mathrm{train}$ for a train of $N$ pulses may be decomposed into a product of
propagators due to individual pulses $U_{p}$: $U_\mathrm{train}=U_{N}...U_{2}U_{1}$.
The pre-train ($t=t_0^{-}$) and post-train ($t=t_0^{+}+NT$) values of the density
matrix are related as $\rho\left(
t_0^{+}+NT\right)=U_\mathrm{train}\,\rho\left(  t_0^{-}\right) \,  U_\mathrm{train}^{\dagger}.$
The propagator across the
$p^{\text{th}}$ pulse may be found analytically~\cite{IliAhmDer11}:
$U_{p}=\boldsymbol{I}\cos\left(  \theta/2\right) +i\boldsymbol{\sigma}_{p}\sin\left(  \theta/2\right)  $, with
$ \boldsymbol{\sigma}_{p} =\boldsymbol{\sigma}_{x}\, \cos\eta_{L,R}(t_p)-
\boldsymbol{\sigma}_{y}\,  \sin\eta_{L,R}(t_p)$, where $\boldsymbol{\sigma}_{x,y}$ are the Pauli matrices.
Here $t_p$ is the arrival time of the $p$-th pulse and $L$ (left) and $R$ (right) label the co- and counter-propagating $\pi$-trains in Fig.~\ref{Fig:Setup}(a). The phase $\eta_{L,R}(t)=\delta \, t\mp k_{c} z(t)-\Phi^{L,R}(t)$ is the cumulative phase of the laser field experienced by the moving molecule, $z(t)\approx v t$. Focusing on a target velocity $v_0$, we may redefine $\Phi^{L,R}(t) = \delta t \mp k_c z_0(t) + \Phi_c^{L,R} (t)$, where $z_0(t)$ is the spacial coordinate of the molecules which initially were at the center of velocity distribution,  $\Phi_c^{L,R} (t)$ is the control phase that we will tune to optimize the cooling process. Then
\begin{equation}
\eta_{L,R}(t)=\mp k_c (z(t)-z_0(t))  -\Phi_c^{L,R}(t) \, .
\label{Eq:etaWithControlPhase}
\end{equation}



At this point we have a prescription for evolving the density matrix over time. The last needed ingredient is the expression for the mechanical momentum transferred to the two-level system.
The fractional momentum kick due to a single train may be expressed in terms of the excited state population difference at the
end and at the beginning of the train
$
-{\Delta p_\mathrm{train}}/{p_{r}}=\rho_{ee}\left(  t_0^{+}+NT\right)-\rho
_{ee}\left(  t_0^{-}\right)$ \cite{IliAhmDer11}.

For the elementary cycle
\begin{equation}
\frac{\Delta p_{\mathrm{cycle}}}{p_{r}}=\rho_{ee}\left( t_0^{+}+2NT
\right) + \rho_{ee}\left(t_0^{-}\right)  -2\rho_{ee}\left(  t_0^{+}+NT\right) \, .
\label{Eq:dpCycle}
\end{equation}

\begin{figure}[h]
\begin{center}
\includegraphics[width=3.5 in]{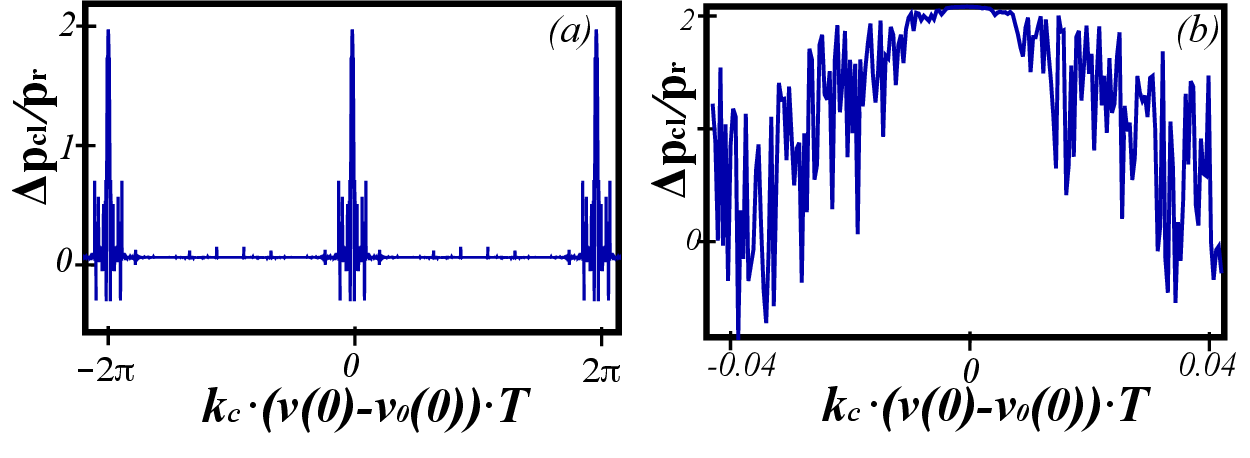}
\end{center}
\caption
{Dependence of the fractional momentum kick per cooling cycle on the Doppler defect phase.
Here $N_c=500$ and $\pi$-trains are composed of four $\pi/4$ pulses.
The overview plot in panel (a) was generated while the velocities were kept constant and control phases $\Phi_c^{L,R}=0$.
Panel (b) zooms in onto the central tooth of panel (a). 
\label{Fig:Zub}}
\end{figure}

Based on the described formalism we developed a Mathematica package to simulate stimulated cooling by coherent pulse trains. We start by discussing our computed dependence of momentum kick per cycle on the Doppler ``defect'' phase
$k_c (v-v_0(t))T$, shown in Fig.~\ref{Fig:Zub}(a). First of all,
$\Delta p$ is a periodic function, reflecting the underlying periodicity of the FC spectrum.
The transferred momentum spikes at values of the phase which are multiples of $2 \pi$. At these points the maximum transferred momentum  is limited to twice the
recoil momentum, as expected from Eq.~(\ref{Eq:dpCycle}). This happens only when during the cycle the system starts and ends in the ground state, with a full transfer of the population to the excited state by the first train, i.e., when the $\pi$-train conditions
are satisfied.
Qualitatively, at such values of the phase the probability amplitudes transferred by subsequent pulses interfere constructively. 



The detailed profile of the momentum transfer shown in Fig.~\ref{Fig:Zub}(b) depends on the number of pulses inside $\pi$-trains and the number of cycles. For a fixed number of pulses the calculated profile  displays a very complicated substructure  resulting from intricate interferences of probability amplitudes driven by the multitude of pulses. In accord with the time-frequency uncertainty principle, as the number of pulses grows larger, the peaks become narrower effectively reducing the capture velocity range to zero. Because of this effect,
it is natural to wonder how to increase the number of decelerated molecules at the end of the process. This goal can be attained by imposing phase relation between the last and the first pulses of two subsequent $\pi$-trains:
\begin{equation}
\Phi_c^{R,L}(t_{M+1, 1})=-\Phi_c^{L,R}(t_{M,N}) + \pi \,. \label{Eq:phaseswitch}
\end{equation}
Here $t_{M,n}$ is the arrival time of the $n$-th pulse of the $M$-th train.
The origin of the phase relation (\ref{Eq:phaseswitch}) may be readily understood using the Bloch sphere visualization method.

\begin{figure}[h]
\begin{center}
\includegraphics*[width=3.5in]{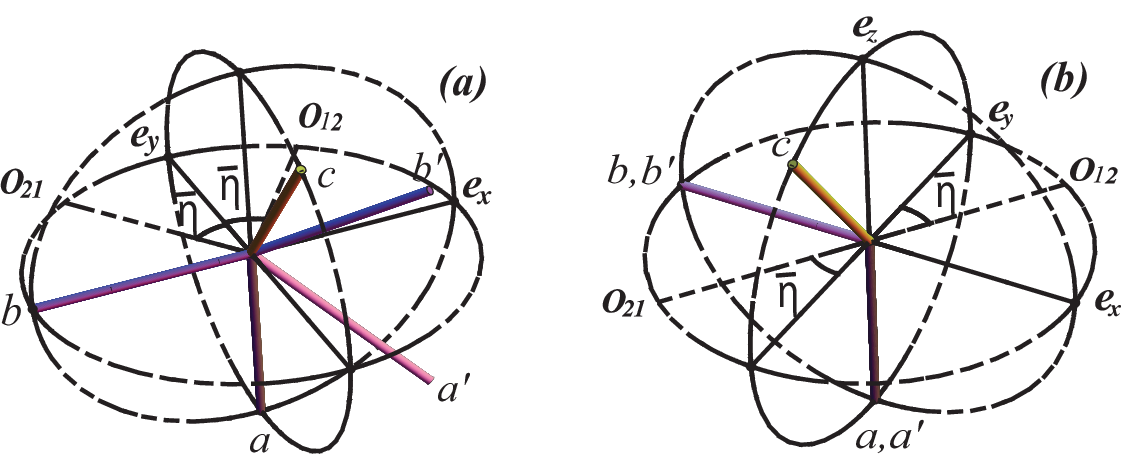}
\end{center}
\caption
{\label{Fig:BlochSphere}
(Color online)
The dynamics of a two-level system driven by a pair of $\pi$-trains visualized on the Bloch sphere. Each train is composed of two $\pi/2$ pulses. Letters a, b, and c (c',b',a') label sequential positions of the Bloch vector during the right (left) train. The system starts in the ground state (vector ``a''). Panels (a) and (b) differ by control phases $\Phi_c^{L,R}$: (a) has no phase correction between the trains and (b) has the correction which reverses the direction of the precession axis; as a result the state vectors returns to the purely ground state at the end of the cycle.}
\end{figure}

This method maps the quantum state vector of a two-level system onto a surface of a sphere. The south/north poles correspond to the pure ground/excited states, respectively.
Other points are uniquely associated with various superposition states.
The interaction with $n$-th pulse of the $M$-th train leads to an instantaneous clockwise rotation of the Bloch vector about the axis $\mathbf{O}_{M,n}$ on angle equal to the pulse area $\theta$.  The $\mathbf{O}_{M,n}$ direction  is determined  by rotating the y-axis about the x-axis on angle $\eta_{L,R}(t_{M,n})$,(see the eq. (3) in the main text), the cumulative phase of the laser field as seen by the moving molecule. The axes $\mathbf{O}_{M,n}$ lie in the x-y plane and their directions can be controlled by choosing the $\Phi_c^{R,L}$ phases.

To begin with, assume that $v=v_0$ and the control phase $\Phi_c=0$. Then the phases $\eta_{L,R}=0$ and the
rotation axes $\mathbf{O}_{M,n}$ coincide with the y-axis. The right  $\pi$-train rotates the Bloch vector from the
south to the north pole of the sphere and the left  $\pi$-train brings it back to the south pole.
If the molecular $v$ is somewhat different from $v_0$, the phase  $\eta_{L,R}=\mp k_c(v-v_0)T$ is no longer zero. This leads to ``skewed'' rotation axes (see Fig.~\ref{Fig:BlochSphere}(a)) and at the end of the train the Bloch vector,  marked by the letter ``c'' in \ref{Fig:BlochSphere} (a), no longer points to a pure excited state. The phase mismatch will accumulate further due to the interaction with the left train.
 This results in momentum kick per cycle much smaller than $2p_c$ or even of the opposite sign.

The phase mismatch problem can be largely mitigated by flipping the direction of the rotation axis for the subsequent $\pi$-train: $\mathbf{O}_{M+1,1}=-\mathbf{O}_{M,N}$ (this translates into the Eq.~(\ref{Eq:phaseswitch})).
Then for $v=v_0$ the Bloch vector when acted upon by the right train exactly retraces the ``left-train'' trajectory in the opposite direction. For $v\neq v_0$, the retracing will not be perfect, but the phase mismatch will be less pronounced compared to the ``uncorrected'' scenario.

In addition to increasing the velocity capture range, phase relation~(\ref{Eq:phaseswitch}) makes our scheme insensitive to laser intensity fluctuations. Laser intensity fluctuations  or laser intensity  spatial variation across the molecular beam can cause the total pulse area of the train to differ from $\pi$.  Indeed, with the phase reversal~(\ref{Eq:phaseswitch})  if the absolute values  of pulse areas of the left and right trains are the same, the Bloch vector in Fig.~\ref{Fig:BlochSphere} exactly retraces the original (driven by the left train) trajectory ending up in the ground state, independent of the value of the pulse area.  Essentially phase relation~(\ref{Eq:phaseswitch}) forces the total pulse area of the two counter-propagating trains to be zero. While our paper was being published, similar mechanism has been elucidated for bichromatic optical forces~\cite{GalAldEyl13}. During the preparation of this manuscript, we became aware of the work of \citet{PhysRevA.89.023425}, which also proposes using stimulated emission to slow molecular beams. \citet{PhysRevA.89.023425} propose solving the problem of incomplete $\pi$ pulses using adiabatic rapid passage, rather than phase reversals as in the current work.

As the molecules slow down, the phases have to track $v_0(t)$ as proposed in Ref.~\cite{IliAhmDer11};
a typical rate of phase-tuning is $d\Phi/dt \approx p_r^2/(\hbar m)/N$, where $m$ is the molecular mass.
Experimentally the required phase-tuning can be attained with electro-optical modulators.

\begin{figure}[h]
\begin{center}
\includegraphics*[width=3.5in]{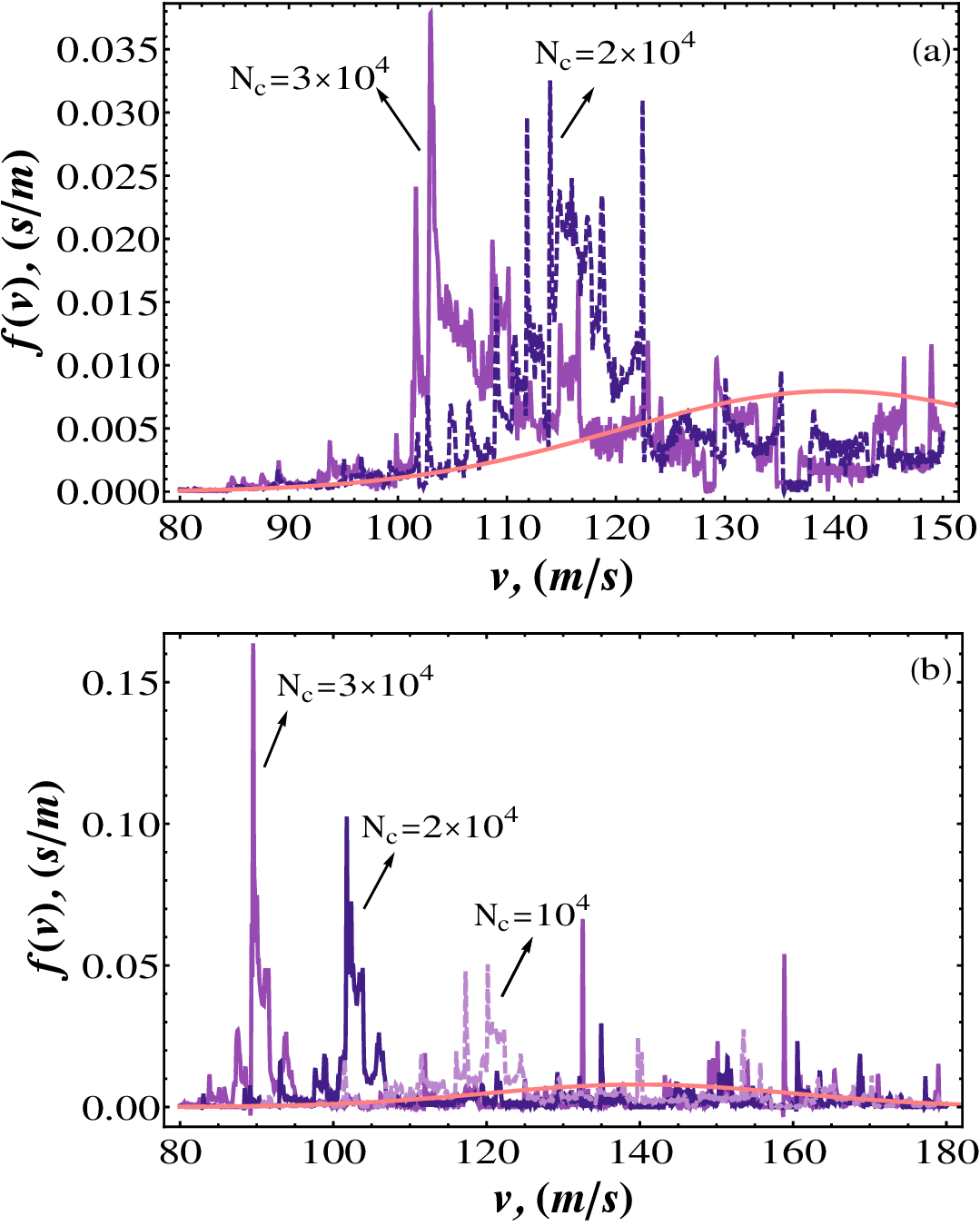}
\end{center}
\caption
{(Color online) Time evolution of velocity distribution of the ensemble of decelerated diatomic molecules LiCl.  All molecules are initially thermally distributed over the rotational states of ground vibrational state of the ground electronic state.
Initial forward velocity  distribution of cryogenic molecular beam is modeled by the gaussian with center at $140\, \mathrm{ m/s}$ and half-width equal to $20\, \mathrm{ m/s}$. Single pulse area $\theta=\pi/4$, pulse repetition period $T=1$ ns. Each cycle takes 8 ns,
and  $N_c=10^4$ corresponds to 80 $\mu$s.
(a) Velocity distributions for the stimulated cooling scheme with Doppler phase tracking and train-to-train phase correction.
(b) The same as in (a) but with additional population reset  applied every 10,000 cycles. \label{Fig:DistroEvolution}
}
\end{figure}

In Fig.~\ref{Fig:DistroEvolution}(a) we investigate the  time evolution of the velocity distribution for LiCl molecules.
We suppose that initial cryogenic molecular beam has a $140\, \mathrm{ m/s}$ forward velocity and a rotational temperature of $\approx 1 K$ \cite{BarShuDem11}.
The forward velocity distribution is modeled as a Gaussian with half-width equal to $20\, \mathrm{ m/s}$.

Initially all molecules are in the ground rovibrational manifold with $(J\leq J_{max})$ and they
enter the interaction volume  at the same moment of time ($t=0$). The laser phase follows the time evolution of $v_0$,
so the  maximum of the optical force is always at $v_0(t)$. To maximize the velocity capture range, the control phase changes as Eq.~(\ref{Eq:phaseswitch}). The strongly peaked force (Fig.~\ref{Fig:Zub}) leads to the compression of velocity distribution near its maximum. However since the force profile narrows with increasing number of cycles $N_c$, the width of the maximum  of velocity distribution starts to decrease too. The area under the maximum also decreases, that means that the effective number of cooled molecules grows smaller. This is caused by a build-up of destructive interferences at the wings of the force profile.

To improve the final number of decelerated molecules one could simply stop the trains and wait until all the molecules radiatively decay into the ground ro-vibrational manifold and then restart the cooling process. This would clear out all the unwanted superpositions. However the required time-scales are too long. Instead, one could employ the much faster radiative decay of the upper electronic states, Fig.~\ref{Fig:Setup}(c).
In this process  additional broad-band laser pulses would drive the transitions: $X(\upsilon,J)\rightarrow A(\upsilon',J'$), with the lower-frequency transitions $X(\upsilon=0,J<J_{\mathrm{max}})\rightarrow A(\upsilon',J')$ filtered out.  Due to the spontaneous decay of the $A(\upsilon',J')$ levels, most of the population after several absorption-radiative decay cycles  gravitates  towards the manifold $X(\upsilon=0, J<J_{max})$ used as a starting point for stimulated cooling.
We simulated such a reset/stimulated cooling scheme in Fig.~\ref{Fig:DistroEvolution}(b). Here the stimulated cooling process is the same as in Fig.~\ref{Fig:DistroEvolution}(a), with the addition of resetting pulses applied every 10,000 cycles.  We observe that
  while the peak of velocity distribution becomes narrower over time, its height increases so that the area (i.e., the number of molecules) increases.

Now we discuss another technical issue: the imperfect polarization of laser pulses.  As an example, consider driving stimulated transitions with 
linearly polarized pulses and denote $\mathcal{A}$ as the degree of 
unwanted circular polarization. Ideally the pulses would drive the $(J,M) - (J+1,M)$ transitions. However, because of the admixture of circular polarization, one would also drive the transitions to the  $(J+1,M-1)$ and $(J+1,M+1)$ magnetic substates. In particular, the states $J+1,M+1$ and $J+1,M-1$ are dark states for linear polarization.
The unwanted fraction of the population transferred due to a single cycle is $\sim \mathcal{A}$ and for $N_c$ cycles we have to require $N_c \mathcal{A} \ll 1 $.  As discussed in Ref.~\cite{PorDerFor04}, reaching the degree of circular polarization $\mathcal{A}= 10^{-6}$ is not considered to be extreme and for our illustrative value of $N_c = 10,000$, this condition is satisfied. Another issue  is the degree of alignment of linear polarizations of two counter-propagating pulse trains. Suppose the two polarization vectors make an angle $\alpha_l$. 
Then the fraction of population transferred into unwanted state is $\sim \alpha_l N_c$ and we have to require $\alpha_l N_c \ll 1$.   Reaching $\alpha_l \sim 10^{-3}$ is considered easy and reducing this angle may require adjusting wave plates using feedback from molecular beam.  
To relax the constraints on   $\mathcal{A}$ and $\alpha_l$ one may reduce the number of cycles before 
the population reset is required.

So far we neglected fine (or hyperfine) structure of molecular states. These could complicate the analysis as the pulses could couple several levels. Still one could isolate two-level transitions using chiral pulse trains. For example,
for the singlet ground state diatomic molecule one could use the $\sigma^+$  ($\sigma^-$ ) circularly-polarized trains. Then a subset of two-level transitions ($\upsilon=0,J,F=I+J,M_F=I+J$)$\leftrightarrows$($\upsilon=1,J'=J+1,F'=I+J+1,M_{F'}=I+J+1$) becomes isolated.

Transverse losses play a role in any deceleration scheme. Traditional Stark decelerators typically suffer significant losses at low velocities due to transverse motion \cite{PhysRevA.73.023401,EuPhysJD.48.197}. These losses can be ameliorated by designs that provide continuous transverse confinement \cite{PhysRevA.81.051401}. Schemes for laser deceleration of atomic beams, such as the Zeeman decelerator, typically exhibit an increase in their angular divergence for two reasons: a decrease in their longitudinal velocity and an increase in their transverse temperature due to spontaneous emission events \cite{PhysRevLett.48.596}. The scheme detailed in this work will exhibit an increase in angular divergence for the former reason, but not the latter, as stimulated emission will leave the transverse velocities unchanged. Possible extensions to our technique could use either transverse cooling or transverse confinement to reduce or eliminate this transverse divergence.

We have demonstrated the method of stimulated decelerating of the ensembles of diatomic molecules by the coherent trains of ultrashort shaped laser pulses. The slowing is based on the stimulated transitions driven within each of the several pairs of rovibrational states in the ground electronic potential.  Since all the transitions are within the ground electronic potential (where the lifetimes of the levels are long), the radiative decay induced  lost of the population to the unwanted states is avoided. The resulting scattering force is based on the stimulated transitions and is  not limited by the small radiative decay rates of the involved levels.  Using the pulse shaping allows one to match the positions of individual FC teeth with the frequencies of chosen molecular transitions.  Manipulating the carrier envelope phase offset between  the subsequent pulses allows to maintain the condition of the resonance for a given velocity group of the ensemble and makes the scheme robust to laser intensity variations.  Manipulating the intensities of individual FC modes allows to keep the pulse area constant for the different rovibrational  transitions  compensating the difference of corresponding dipole matrix elements. The evolution of the velocity distribution of cryogenic molecular beam decelerated according to the proposed scheme has been calculated.  The scattering rate depends on the momentum kick exerted on the molecule during the single working cycle, which maximum is determined as $2p_r$, where the recoil momentum  is $p_r\approx \hbar k_c$.  As an example for the LiH molecule $v_r\approx 0.001 $ m/s. That is about $N_c\sim 10^5$  pairs of $\pi$-trains is needed to decelerate the ensemble of molecules with initial central velocity $v_c\sim 100$ m/s.

\acknowledgments
This work  was supported in part by the US National Science Foundation.

%

\end{document}